\documentstyle[pre,aps,twocolumn,epsf]{revtex}
\begin{document}
\draft
\title{Percolation transition and the onset of non exponential relaxation
in fully frustrated models}
\author{Annalisa Fierro$^{1,2}$, Giancarlo Franzese$^{1,2}$,
        Antonio de Candia$^{1,2}$, Antonio Coniglio$^{1,2}$}
\address{$^1$ Universit\`a Degli Studi di Napoli ``Federico II'',
              Dipartimento di Scienze Fisiche,}
\address{Mostra d'Oltremare, Pad. 19, 80125 Napoli, Italy}
\address{$^2$INFM, Unit\`a di Napoli, Napoli, Italy}
\maketitle
\begin{abstract}
We numerically study the dynamical properties of fully frustrated models in 2 
and 3 dimensions.
The results obtained support the hypothesis that the percolation transition
of the Kasteleyn-Fortuin clusters corresponds to the onset
of stretched exponential autocorrelation functions in systems without disorder.
This dynamical behavior may be due to the ``large scale'' effects of 
frustration, present below the percolation threshold.
Moreover these results are consistent with the picture suggested by Campbell 
{\em et al.} in space of configurations.
\end{abstract}
%
%%%%%%%%%%%%%%%%%%%%%%%%%%%%%%%%%%%%%%%%%%%%%%%%%%%%%%%%%%%%%%%%%%%%%
%
\section{Introduction}
At low temperature spin glasses (SGs) undergo a transition characterized by the 
divergence of the nonlinear susceptibility.
Moreover the relaxation functions of the system become non exponential already
at temperatures higher than the transition temperature $T_{sg}$.

This behavior has been observed in canonical metallic and insulating spin 
glasses, investigated by neutron and hyperfine techniques
\cite{ref_exp}.

In the Ising SG model, studied with spin-flip Monte Carlo dynamics, 
both in 2 dimensions ($2d$) \cite{ref_mcmillan} and  $3d$  \cite{ref_ogielski}, 
non exponential relaxation functions have been observed below some temperature
$T^*$ higher than $T_{sg}$. Moreover in the $3d$ system Ogielski 
\cite{ref_ogielski} observed that the long time
regime of the relaxation functions is well approximated by the following 
function
\begin{equation}
f(t)=f_0~t^{-x}~\exp[-(t/\tau)^\beta].
\label {eq_ogielski}
\end{equation}
Fitting the data with this function he obtained that the onset of non 
exponential relaxation is consistent with the Griffiths temperature $T_G$, 
that coincides with the critical temperature of the ferromagnetic model.

This result supports the argument suggested by Randeria {\em et al.} 
\cite{ref_randeria},
recently verified by more rigorous analysis in Ref.\cite{ref_cesi} and by 
numerical simulations on a generalization of the SG model in 
Ref.\cite{ref_fierro,FC_PRE}.
According to these analysis in the SG the onset $T^*$
of the non exponential relaxation should be greater than or equal to the 
Griffiths temperature $T_G$. This behavior is caused by the existence of 
unfrustrated 
ferromagnetic-type clusters of interactions, the same that are responsible for
the Griffiths singularity \cite{ref_griffiths}.
The presence of non exponential relaxation in this approach is therefore
a direct consequence of the quenched disorder.

Another mechanism leading to non exponential relaxation in frustrated 
systems, such as SG,
has been suggested by several authors 
\cite{ref_campbell,ref_silvia,ref_glotzer}.
According to these arguments the onset $T^*$ of non exponential relaxation
is greater than or equal to the percolation transition $T_p$ of the 
Kasteleyn-Fortuin  and Coniglio-Klein (KFCK) clusters 
\cite{ref_kasteleyn,ref_cklein}. However, in frustrated systems with disorder 
$T_p$ is less than but close to $T_G$, therefore its eventual effects are 
hidden by those related to $T_G$. 

A way to verify if percolation mechanisms can play a role in the dynamical 
transition of frustrated systems is to consider frustrated models without 
disorder where the Griffiths phase is not defined. In particular, we have 
considered fully frustrated (FF) 
spin systems 
\cite{ref_villain}
%,ref_forgacs,
where ferromagnetic and 
antiferromagnetic interactions are 
distributed in a regular way on the lattice, in such a way that 
no unfrustrated region (no Griffiths phase) exists, but 
the percolation temperature of KFCK clusters is still defined.

In a previous paper \cite{ref_fierro} we have studied the $2d$ FF
Ising model. We found numerically that the model 
exhibits a non exponential relaxation below the  percolation
temperature $T_p$ of the KFCK clusters. 
Moreover the long time regime of these functions is well approximated  
by a Kohlrausch-Williams-Watts function, also known as
``stretched exponential'',
\begin{equation}
f(t)=f_0\exp[-(t/\tau)^\beta].
\label{eq_stretched}
\end{equation}

In this paper we analyse, with conventional spin flip,
the dynamical behavior of FF Ising model in $3d$ and in $2d$
using better statistics finding that $T^*$ is numerically consistent with 
$T_p$. 

To clarify the role of percolation 
we study also the $3d$ $q$-bond FF percolation model. 
For $q=2$ this model is obtained applying the KFCK cluster formalism to 
the FF Ising model
(see Sect. \ref{sec_model}). We simulate it using the ``bond flip" dynamics 
\cite{ref_fierro}. In this way the percolation properties of the model are 
stressed and the appearing of non exponential relaxation functions 
at $T_p$ are more evident.

In both these cases we find that the relaxation functions exhibit an 
exponential long time behavior at high temperatures.
Below the percolation temperature $T_p$ of the KFCK clusters, 
that is higher than the transition temperature $T_c$ of the model, 
the long time regime of the relaxation 
functions becomes non exponential and is well approximated by a stretched 
exponential.

Our results are consistent with the picture suggested by Campbell {\em et al.}
\cite{ref_campbell} in the space of configurations and can be interpreted 
considering that $T_p$ corresponds to a thermodynamic transition in a 
generalized frustrated model \cite{FC_PRE}.

In Sect. \ref{sec_model} we present the ``$q$-bond frustrated percolation''
model, and in Sect. \ref{sec_statics} we study the percolation properties
of this model on a FF cubic lattice.
We find that the percolation transition is in the same universality
class of the $q/2$-state ferromagnetic Potts model confirming the results 
obtained in the disordered version of the model in $2d$ \cite{FC_PRE}.

In Sect. \ref{sec_sflip} we study the FF Ising model dynamical
properties with conventional spin flip, and in Sect. \ref{sec_relaxation}
we present the relaxation functions obtained simulating the FF 
$q$-bond percolation model for $q=2$, with the ``bond flip''
dynamics.

In Sect. \ref{campbell} we show the connection with the Campbell scenario 
\cite{ref_campbell} and in Sect. \ref{conclusions} we give the conclusions.
\section{The ``\lowercase{$q$}-bond frustrated percolation'' model}
\label{sec_model}
\begin{figure}
\begin{center}
\mbox{ \epsfxsize=3cm \epsffile{ 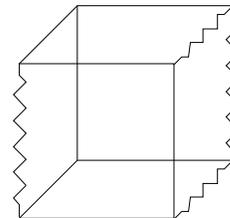 } }
\caption{Distribution of interactions for the FF model.
Straight lines and wavy lines correspond, respectively, to
$\epsilon_{ij}=1$ and $\epsilon_{ij}=-1$.}
\end{center}
\label{fig_fully}
\end{figure}

The FF Ising spin model is defined by the Hamiltonian
\begin{equation}
{\cal H}=-J\sum_{\langle{ij}\rangle}(\epsilon_{ij}S_{i}S_{j}-1),
\label{eq_Ising}
\end{equation}
where $\epsilon_{ij}$ are quenched variables which assume the
values $\pm 1$. The ferromagnetic and antiferromagnetic interactions are
distributed in a regular way on the lattice (see Fig. \ref{fig_fully}).

Using the KFCK cluster formalism for
frustrated spin Hamiltonians \cite{ref_diliberto},
it is possible to show that the partition function of the model Hamiltonian
in Eq. (\ref{eq_Ising})
is given by
\begin{equation}
Z={\sum_C}^\ast e^{\mu n(C)/k_B T} q^{N(C)},
\label{eq_partition}
\end{equation}
where $q=2$ is the multiplicity of the spins, $k_B$ is the Boltzmann constant,
$\mu=k_BT\ln(e^{qJ/k_B T}-1)$,
$n(C)$ and $N(C)$ are respectively the number of bonds and the number of
clusters in the bond configuration $C$. The summation $\sum_C^\ast$ extends
over all the bond configurations that do not contain a ``frustrated loop'',
that is a closed path of bonds which contains  an odd number of
antiferromagnetic interactions.
Note that there is only one parameter in the model, namely the temperature
$T$, ranging from $0$ to $\infty$. The parameter $\mu$,
that can assume positive
or negative values, plays the role of a chemical potential.

Varying $q$ we obtain an entire class of models differing by the
``multiplicity'' of the spins, which we call the 
$q$-bond FF percolation model.
More precisely, for a general value of $q$,
the model can be obtained from a Hamiltonian \cite{ref_vittorio}
\begin{equation}
{\cal H}=-sJ\sum_{\langle{ij}\rangle}
[(\epsilon_{ij}S_{i}S_{j}+1)\delta_{\sigma_{i}\sigma_{j}}-2],
\label{Hs}
\end{equation}
in which every site carries two types of spin, namely an Ising spin and a Potts
spin $\sigma_i=1,\ldots,s$ with $s=q/2$.
For $q=1$ the factor $q^{N(C)}$ disappears
from Eq. (\ref{eq_partition}), and we obtain a simpler model in which
the bonds are randomly distributed under the conditions that the bond 
configurations do not contain a frustrated loop.
For $q\to 0$ we recover the tree percolation,
in which all loops are forbidden, be they frustrated or not
\cite{ref_wu}.

When all the interactions are positive (i.e. $\epsilon_{ij}=1$)
the sum in Eq. (\ref{eq_partition}) contains all bond configurations
without any restriction. In this case the partition function coincides 
with the partition function of the ferromagnetic $q$-state Potts model, 
which in the limit $q=1$ gives the random bond percolation \cite{ref_wu}.

 From renormalization group \cite{ref_pezzella}, mean field \cite{DLP} 
and numerical results 
\cite{FC_PRE,ref_adc}  we expect that the
model in Eq. (\ref{Hs}) exhibits two critical points:
the first at a temperature $T_p(q)$, corresponding to the percolation
of the bonds on the lattice, in the same universality class of the
ferromagnetic $q/2$-state Potts model; 
the other at a lower temperature $T_c(q)$, in the
same universality class of the FF Ising model. 
\section{Static properties}
\label{sec_statics}
In this Section we analyse the percolation properties of the model defined
by Eq. (\ref{Hs}) for $q=2$, on a FF cubic
lattice.

After preliminary runs with spin flip dynamics on systems with lattice sizes
$L=10,20$, and with statistics of $5\times10^3$ thermalization steps
and $5\times10^6$ acquisition steps, we found that the percolation 
transition
occurs well above the critical temperature $T_c=1.35$ \cite{ref_deep}
(in the following the 
temperatures will be given in $J/k_B$ units). 
Then, we have simulated the model for $L=30\div80$, 
by the Swendsen and Wang cluster dynamics \cite{ref_SW87}, that turns 
out to be very efficient for the temperature regime of interest, allowing
to consider only $5\times10^4$ acquisition steps.

At every step we have evaluated the percolation probability
\begin{equation}
P=1-\sum_{s} s n_s,
\end{equation}
and the mean cluster size 
\begin{equation}
S=\sum_{s} s^2n_s,
\end{equation}
where $n_s$ is the density of finite clusters of size $s$.

Around the percolation temperature, the averaged
quantities $P(T)$ and
$S(T)$, for different values of the lattice size $L$,
should obey to the finite size scaling \cite{ref_aharony}
\begin{mathletters}
\begin{equation}
P(T) = L^{-\beta/\nu}F_P[L^{1/\nu}(T-T_p)],
\end{equation}
\begin{equation}
S(T) = L^{\gamma/\nu}F_S[L^{1/\nu}(T-T_p)],
\end{equation}
\end{mathletters}
where $\beta$, $\gamma$ and $\nu$ are critical exponents, $F_P(x)$ and 
$F_S(x)$ are universal functions of an adimensional quantity $x$.

\begin{figure}
\begin{center}
\mbox{ \epsfxsize=6cm \epsffile{ 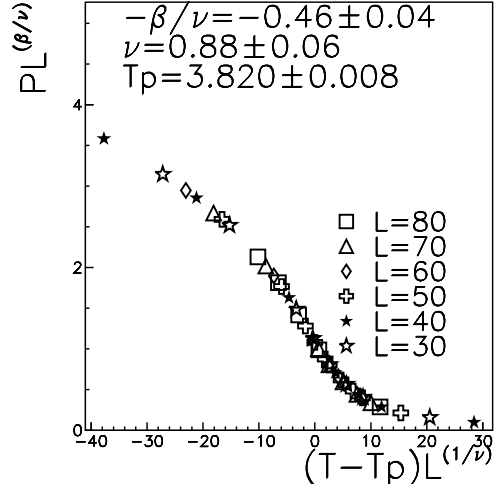 } }
\mbox{ \epsfxsize=6cm \epsffile{ 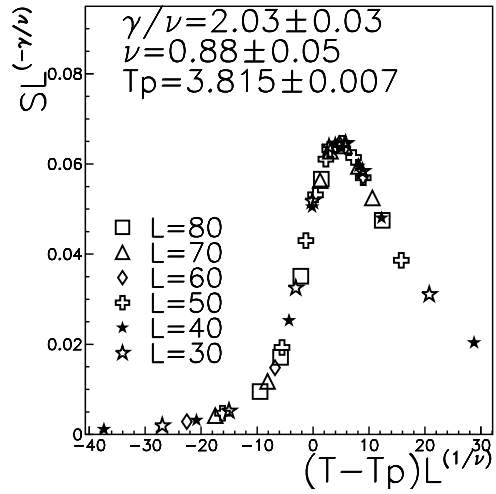 } }
\caption{Finite size scaling of (a) $P(T)$ and (b) $S(T)$,
for the $q=2$ model, and for lattice sizes $L=30$,40, 50, 60, 70, 80.}
\end{center}
\label{fig_statics}
\end{figure}

Standard scaling analysis results are summarized in Fig. \ref{fig_statics}. 
We obtained 
$T_p=3.817\pm 0.005$, 
$\nu=0.88\pm 0.06$, $\beta/\nu=0.46\pm 0.04$ and $\gamma/\nu=2.03\pm 0.03$. 
The values of the critical exponents coincides, within the errors, with the 
random bond percolation exponents \cite{ref_aharony}.

As we expect, the $q=$2-bond frustrated percolation model is in the
same universality class of the $q/2=$1-state ferromagnetic Potts model.

\section{The relaxation functions of the fully frustrated Ising spin model}
\label{sec_sflip}

In this Section we present our results in the study of the
FF Ising model, defined by the Hamiltonian in Eq. (\ref{Hs})
for $q=2$,
simulated by spin flip dynamics.

For each temperature $T$, 16
different runs were made, varying the random number generator seed,
on a FF cubic lattice of size $L=30$. We took about $10^4$ steps
for thermalization, and about $10^5$ steps
for acquisition, calculating at each step the energy $E(t)$.
The relaxation function of the energy is defined as
\begin{equation}
f(t)=\frac{\langle\delta E(t)\delta E(0)\rangle}
{\langle(\delta E)^2\rangle},
\end{equation}
where $\delta E(t)=E(t)-\langle E\rangle$. For each value of
$T$, we averaged the 16 functions calculated and evaluated the
error as a standard deviation of the mean.
Here a unit of time is considered to be one Monte Carlo
step, that is $L^d$ single spin update trials.

\begin{figure}
\begin{center}
\mbox{ \epsfxsize=6cm \epsffile{ 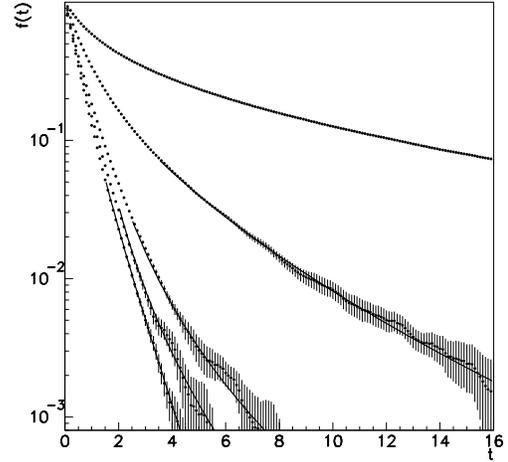 } }
\caption{Relaxation functions $f(t)$ of energy as function of time
$t$ for the $d=3$ FF Ising model, with spin flip dynamics,
lattice size $L=30$,
for temperatures (from left to right) $T=4.0$, 3.5, 3.0, 2.0, 1.5.}
\end{center}
\label{fig_relax_sflip}
\end{figure}

In Fig. \ref{fig_relax_sflip} we show the results for
$T=4.0$, 3.5, 3.0, 2.0, 1.5. 
We observe a two step decay also for high temperatures. 
For all the temperatures we fit the long time tail of the relaxation 
functions with the empirical formula proposed by Ogielski in Eq. 
(\ref {eq_ogielski}).

\begin{figure}
\begin{center}
\mbox{ \epsfxsize=6cm \epsffile{ 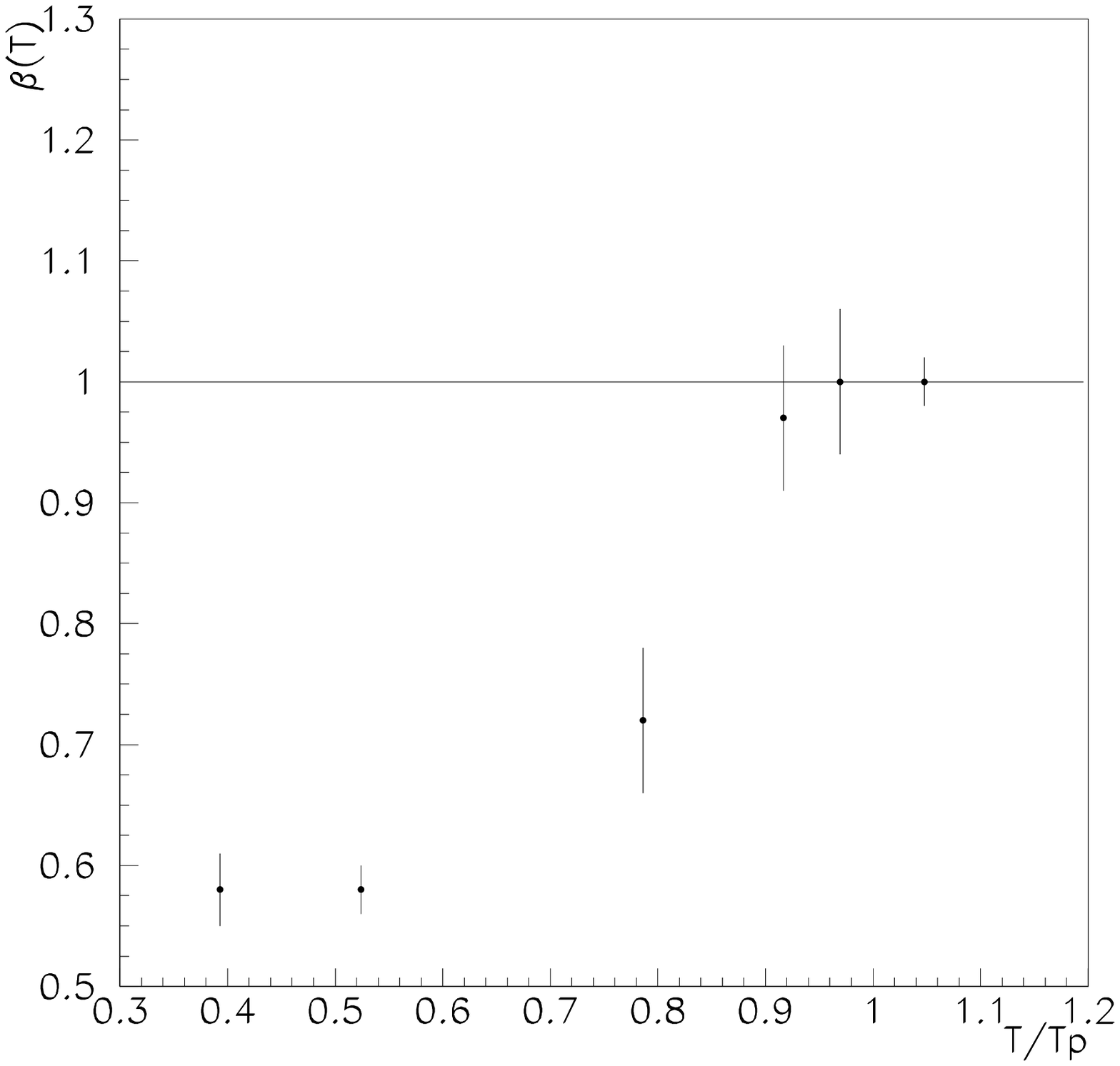 } }
\caption{Stretching exponents $\beta(T)$ as function of $T/T_p$, the ratio
of temperature over percolation temperature, 
for the $d=3$ FF Ising model, with spin flip dynamics,
lattice size $L=30$.}
\end{center}
\label{fig_sflip_betas}
\end{figure}

Temperature dependence of exponents $\beta(T)$ is presented
in Fig. \ref{fig_sflip_betas}.
Note that $\beta(T)$ increases as function of $T$ from the value
$\beta=0.58\pm0.03$ for $T=1.5$ to the value $\beta=1$ for $T= 3.7$, 4.0.
We do not observe any regular behavior in the temperature
dependence of exponent $x(T)$.
We estimated the errors on parameters as the range where we obtain a 
good fit of the relaxation function.  

As we can see in Fig. \ref{fig_sflip_betas}, these results are consistent,
within the errors, with the scenario in which the onset of the stretched 
exponential relaxation coincides with the percolation temperature $T_p=3.817\pm 0.005$ 
(see Sect. \ref{sec_statics}). 

We also simulated the FF Ising model on a square lattice
of size $L=60$.
We calculated the relaxation functions of the energy. Averages were made
over $16$ different random generator seeds, and between $10^5$ and $10^6$
steps for acquisition were taken, after about $10^4$ steps for 
thermalization.

\begin{figure}
\begin{center}
\mbox{ \epsfxsize=6cm \epsffile{ 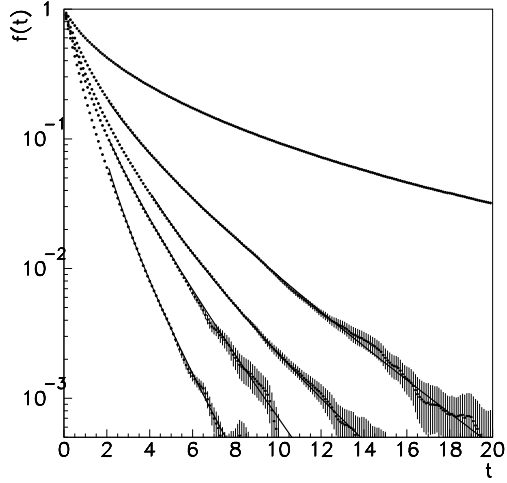 } }
\caption{Relaxation functions $f(t)$ of energy as function of time
$t$ for the $d=2$ FF Ising model, with spin flip dynamics,
lattice size $L=60$,
for temperatures (from left to right) $T=2.5$, 2.0, 1.8, 1.5, 1.0.}
\end{center}
\label{fig_relax_sflip_d2}
\end{figure}

In Fig. \ref{fig_relax_sflip_d2} we show the relaxation functions obtained 
for $T=2.5$, 2.0, 1.8, 1.5, 1.0. 
For all temperatures we fit the long time tail of the 
relaxation  functions with Eq. (\ref{eq_ogielski}).

\begin{figure}
\begin{center}
\mbox{ \epsfxsize=6cm \epsffile{ 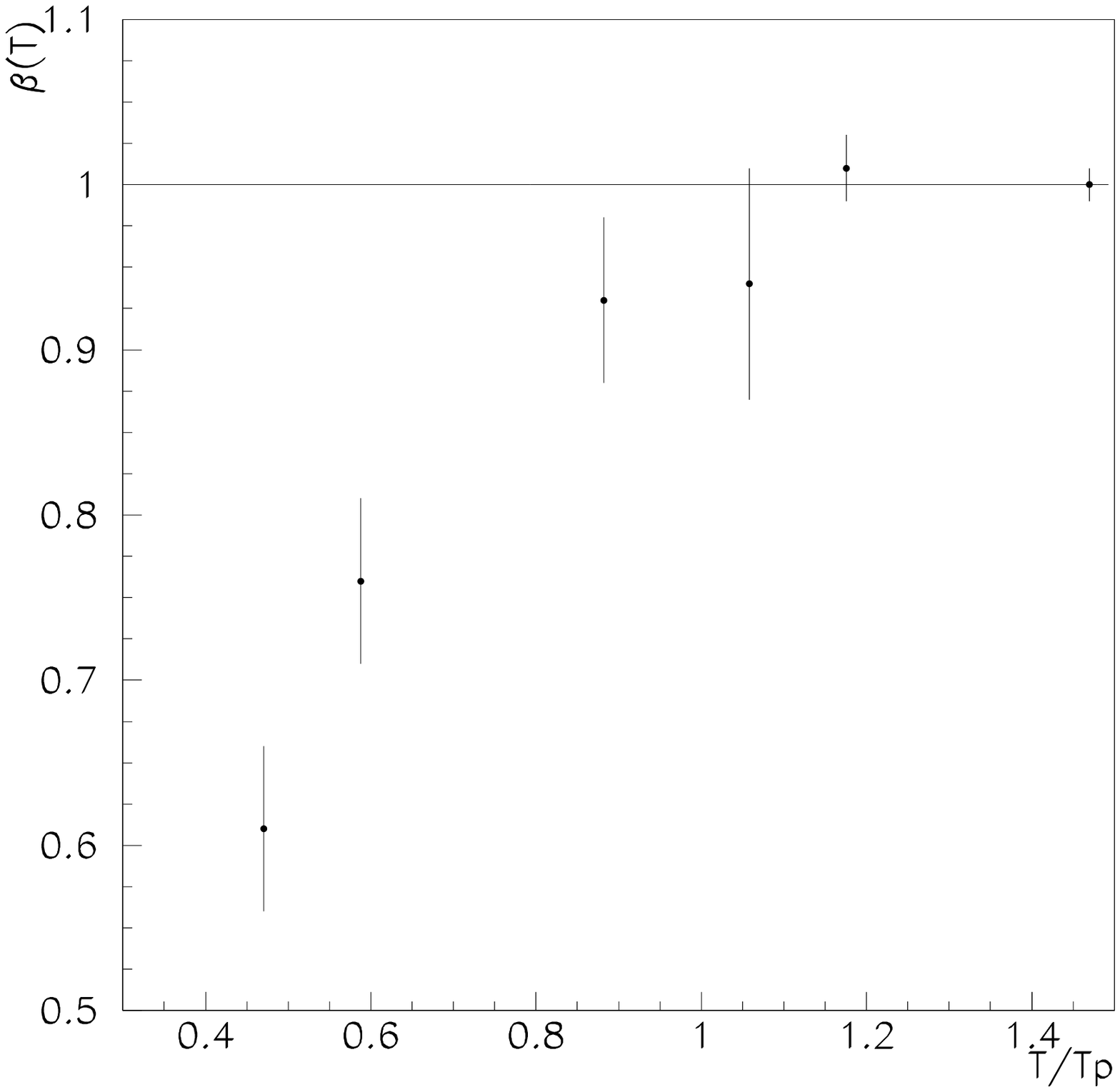 } }
\caption{Stretching exponents $\beta(T)$ as function of $T/T_p$, the ratio
of temperature over percolation temperature, 
for the $d=2$ FF Ising model, with spin flip dynamics,
lattice size $L=60$.}
\end{center}
\label{fig_sflip_betas_d2}
\end{figure}

The temperature dependence of exponents $\beta(T)$ is shown
in Fig. \ref{fig_sflip_betas_d2}.
Note that $\beta(T)$ increases as function of $T$ from the value
$\beta=0.61\pm0.05$ for $T=0.8$ to the value $\beta=1$ for $T\geq 2.0$.

As we can see in Fig. \ref{fig_sflip_betas_d2}, our estimate of the onset of 
the stretched exponential relaxation is also consistent, within the errors, 
with the percolation temperature $T_p=1.701$ \cite{ref_fierro}.

Within the errors the exponent $x(T)$ increases as function of $T$ 
from the value $x=0.4\pm0.2$ for $T=0.8$ to the value $x=1.6\pm 0.4$ 
for $T= 2.5$.
\section{The relaxation functions of the
``\lowercase{$q$}-bond frustrated percolation'' model}
\label{sec_relaxation}
In this Section we analyse the dynamical behavior of the model defined by
Eq. (\ref{eq_partition})
with $q=2$, simulated by the bond flip dynamics \cite{ref_fierro}. 

The dynamics is carried out in the following way: 
at each step we choose at random a particular
edge on the lattice; calculate the probability $P$ of changing its state,
that is of creating a bond if the edge is empty, and of destroying the bond
if the edge is occupied; and, finally, we change the state of the edge
with probability $P$.

For each temperature $T$, 16
different runs were made, varying the random number generator seed,
on a FF cubic lattice of size $L=30$. We took about $10^3$ steps
for thermalization, and between $10^4$ and $10^5$ steps
for acquisition, calculating at each step the density of bonds $\rho(t)$.
The relaxation function of the density of bonds is defined as
\begin{equation}
f(t)=\frac{\langle\delta\rho(t)\delta\rho(0)\rangle}
{\langle(\delta\rho)^2\rangle},
\end{equation}
where $\delta\rho(t)=\rho(t)-\langle\rho\rangle$. For each value of
$T$, we averaged the 16 functions calculated and evaluated the
error as a standard deviation of the mean.
We consider a unit of time
to consist of ${\cal G}\langle\rho\rangle^{-1}$ single update trials,
where ${\cal G}=3L^3$ is the number of edges on the lattice.

\begin{figure}
\begin{center}
\mbox{ \epsfxsize=6cm \epsffile{ 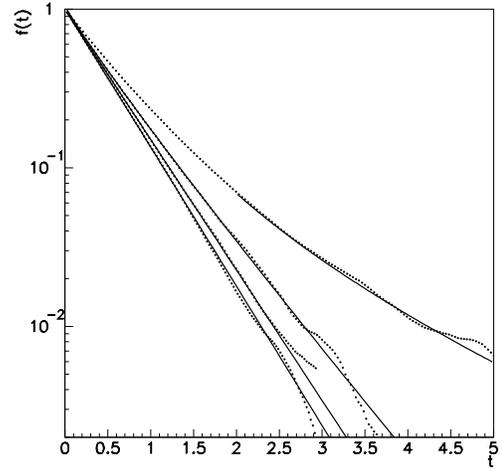 } }
\caption{Relaxation functions $f(t)$ of bond density as function of time
$t$, for the $q=2$ FF bond percolation model, 
on a $d=3$ lattice of size $L=30$,
for temperatures (from left to right) $T=4.0$, 3.5, 3.0, 2.5.}
\end{center}
\label{fig_relax}
\end{figure}

In Fig. \ref{fig_relax} we show  the results obtained for temperatures 
$T=4.0$, 3.5, 3.0, 2.5.
For $T=4.0$, 3.5 we fitted the calculated points with the function
in Eq. (\ref{eq_ogielski}).

The value of $\beta$ extracted from the fit is equal to one 
within the error, and the value of $x$ is zero. 
Thus for these temperatures the relaxation is purely exponential.

\begin{figure}
\begin{center}
\mbox{ \epsfxsize=6cm \epsffile{ 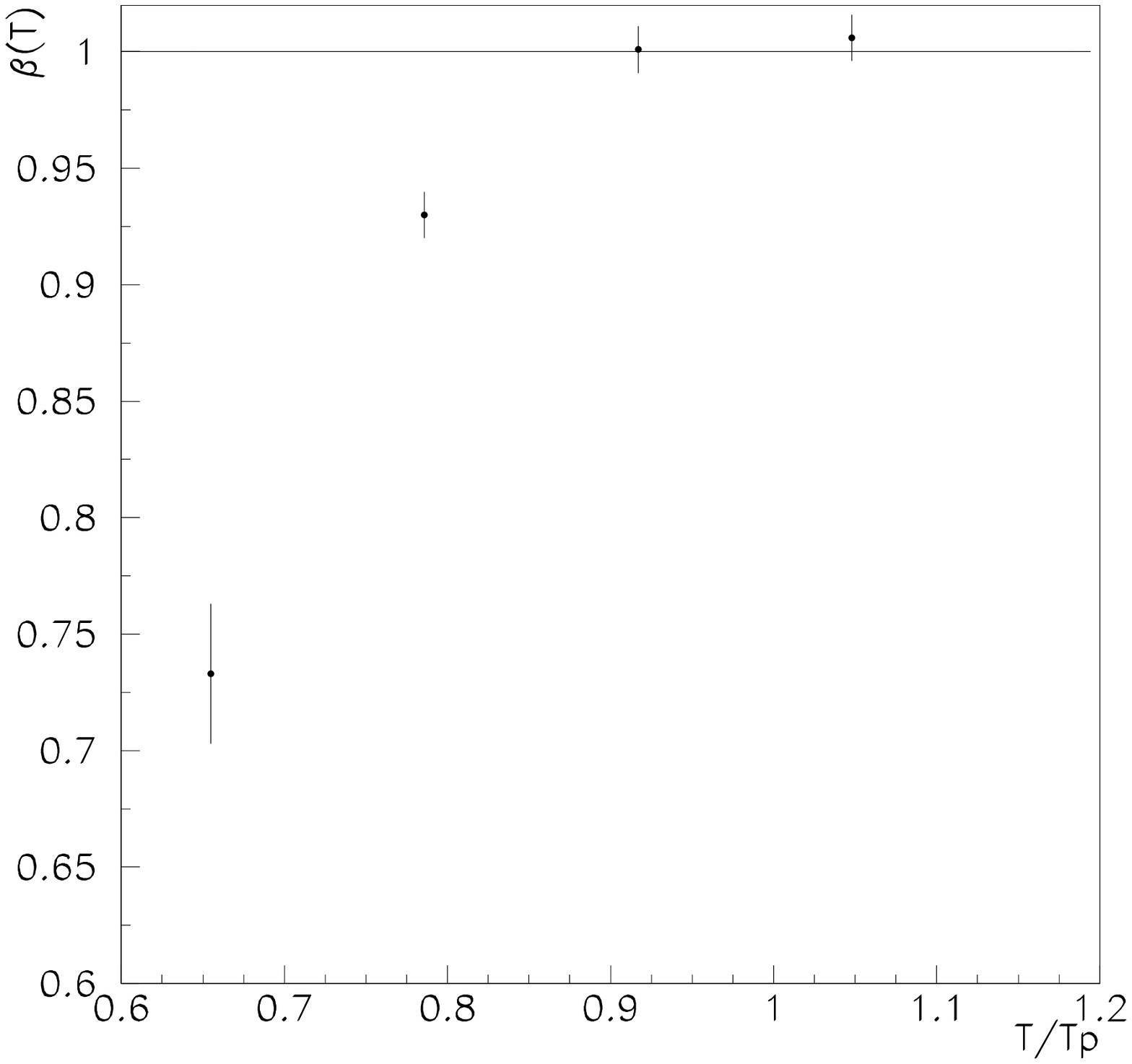 } }
\caption{Stretching exponents $\beta(T)$ as function of $T/T_p$, the ratio
of temperature over percolation temperature, for the $q=2$ FF
bond percolation model, on a $d=3$ lattice of size $L=30$}
\end{center}
\label{fig_beta_1}
\end{figure}

For $T<3.5$ we observe a two step decay, and only the long time regime of the 
relaxation functions could be fitted by Eq. (\ref{eq_ogielski}).
The value of $\beta$ extracted is less than one, showing that stretched 
exponential relaxation has appeared for these temperatures.
In Fig. \ref{fig_beta_1} the values of $\beta(T)$ as function of the
ratio $T/T_p$ are shown, with errors estimation.
The exponent
$x(T)$ becomes non zero only for $T=2.5$,
for this value of temperature we obtain $x=1.1 \pm 0.1$

As we can see in Fig. \ref{fig_beta_1}, our estimate of the onset of 
stretched exponential relaxation is consistent, within the errors,
with the percolation temperature $T_p=3.817\pm 0.005$ 
of the KFCK clusters.

\section{Connection with the random walk picture}
\label{campbell}
In this Section we make a connection 
between our model and 
the random walk picture
of Campbell {\em et al.} \cite{ref_campbell}, which we will briefly 
illustrate. 
Consider an hypercube in a D-dimensional space. Each summit is 
occupied with a probability $p$. On such a dilute lattice,
a random walker is allowed to diffuse, 
like the ``ant"  on a percolating cluster in the de Gennes
picture. The mean square displacement after a time t is given by 

\begin{equation}
r^2(t)\equiv\frac{\langle\sum_{i=1}^D(x_i(t)-x_i(0))^2\rangle}{D},
\label{ref_r2}
\end{equation}
where $D$ is the hypercube dimension, $x$ is a $D$-dimensional vector of 
components $0, 1$, that identify the hypercube $2^D$ summits, and $x(t)$
indicates the ``ant'' position at the time $t$.

Campbell {\em et al.} 
suggest in the Ising SG model that accessible 
region in the space of configurations, 
compact at high temperature, becomes ramified at a temperature $T^*$,
and that a complex space of configurations
 is responsible for the appearing of non 
exponential relaxation. They also suppose that this temperature $T^*$
is the percolation temperature of the KFCK clusters.
The idea is that the diffusive ant mimics quite well the
evolution in the space of configurations in the SG model.

In the study of the random walk on a randomly occupied hypercube
they find that for $p\le p^*$ the function $r^2(t)$ becomes non exponential and
is well approximated by a stretched exponential. But it is not possible
to associate a value of temperature to this probability.

To make the connection between the bond frustrated percolation 
formalism and the random walk picture we introduce the local bond density
autocorrelation function $f(t)$

\begin{equation}
f(t)=\frac
{\sum_{i=1}^{\cal G}\langle b_i(t)b_i(0)\rangle-\langle b_i(t)\rangle^2}
{\sum_{i=1}^{\cal G}\langle b_i(t)\rangle-\langle b_i(t)\rangle^2},
\label{autocorr}
\end{equation}
where ${\cal G}=dL^d$ is the number of lattice edges, $d$ is the lattice 
dimension and $b_i=0$ if 
the i-th bond is missing and $b_i=1$ if the i-th bond is present.
The variables 
$b_i(t)$ are the coordinates in the space of configurations
of the $q-$bond FF 
percolation model, 
which evolve by bond flip dynamics. In analogy with 
the picture proposed by Campbell {\em et al.},
these variables   
can be interpreted as  the coordinates of a walk on 
the summits of a ${\cal G}$-dimensional hypercube.
For a fixed value of the temperature the walk will be confined
in the subspace with density of bonds corresponding to that
temperature\footnote{%
The walk in such subspace is not random, since each step has a weight,
which is a function of the clusters number variation.
If we consider the $q=$1-bond 
frustrated percolation model, the walk would be random.}.

Due to frustration not all configurations are allowed, therefore 
the walk occurs on a dilute space. By changing the temperature the 
space of configurations where the walk is confined 
changes and therefore  
the density of allowed  sites in such a region of the space of configurations 
also changes. 
This is realized in a artificial way in the picture made by Campbell 
{\em et al.}, occupying randomly the hypercube summits
(that represent the accessible states in the space of configurations).
By changing
the temperature, one may reach therefore
a percolation threshold in the space of configurations. 
This would correspond to the breaking of ergodicity.
At higher temperature, however, the space of configurations may become ramified
and stretched exponentials start to appear.

Eq. (\ref{autocorr}) can be related to the distance travelled by 
the random  walk $r(t)$, 
via the relation $r^2(t)=2(\langle\rho\rangle-f(t))$.

We have simulated, by bond flip dynamics,
the $q=2$-bond FF percolation model on a square lattice of size $L=60$.
We have calculated the temperature dependence of the autocorrelation 
function in Eq. (\ref{autocorr}).
We find  an exponential relaxation at high temperatures, while for $T<T^*$
the long time behavior of relaxation functions becomes non exponential and 
is well approximated by a stretched exponential. 

In Tab. \ref{tabella} we show the fit parameters. Our estimate of the onset of 
stretched exponential relaxation functions is also consistent, within the 
errors, with the percolation temperature $T_p=1.701$.

\begin{figure}
\begin{center}
\mbox{ \epsfxsize=6cm \epsffile{ 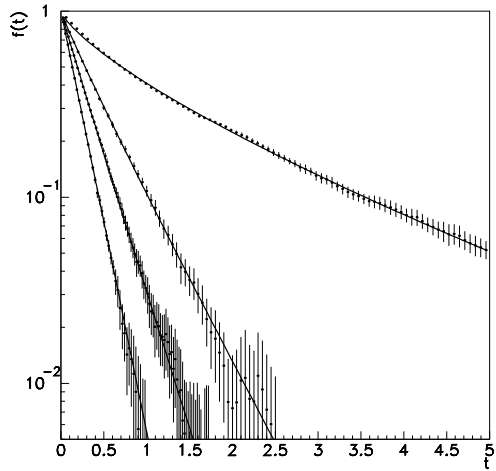 } }
\caption{Autocorrelation functions $f(t)$ as function of time $t$ 
for the $q=2$ FF bond percolation model, 
on a $d=2$ lattice of size $L=60$,
for temperatures (from left to right) $T=3.5$, 2.0, 1.7, 1.3.}
\end{center}
\label{hamming_1}
\end{figure}

\begin{figure}
\begin{center}
\mbox{ \epsfxsize=6cm \epsffile{ 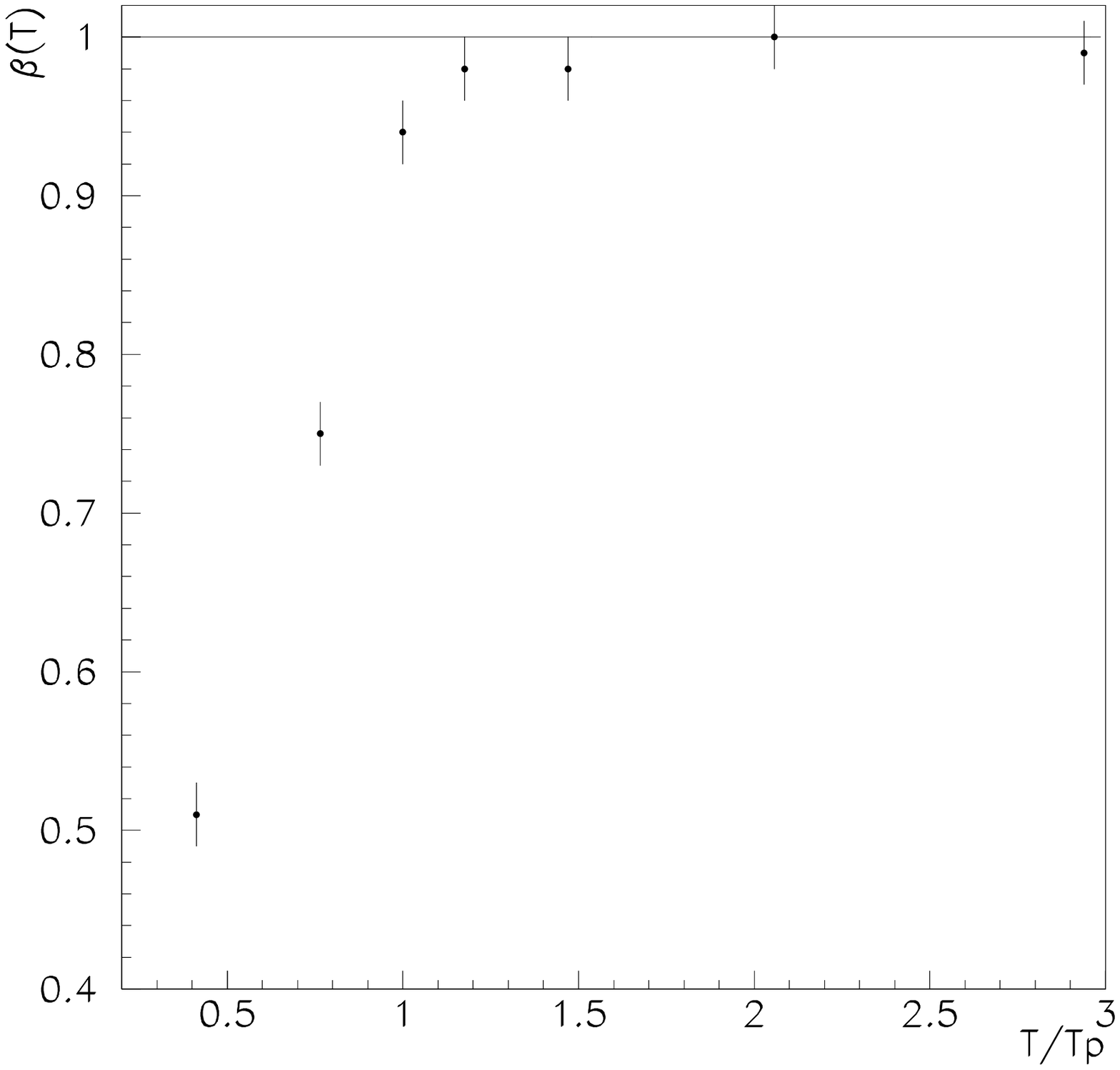 } }
\caption{Stretching exponents $\beta(T)$ as function of $T/T_p$, the ratio
of temperature over percolation temperature, for the $q=2$ FF 
bond percolation model, on a $d=2$ lattice of size $L=60$}
\end{center}
\label{hamming_2}
\end{figure}

In Fig. \ref{hamming_1} we show the functions $f(t)$ for 
temperatures $T=3.5,$ 2.0, 1.7, 1.3,
and in Fig. \ref{hamming_2} the temperature dependence of the exponents 
$\beta(T)$ as a function of $T/T_p$.

We conclude
that it is possible to apply the picture proposed by Campbell {\em et al.}
to the $q$-bond frustrated percolation model.
Furthermore our results are consistent with the hypothesis that the onset of 
non exponential relaxation function coincides with $T_p$.
Note that we cannot exclude numerically 
that stretched exponentials are present even at temperatures higher
than the percolation transition,
with an extremely small amplitude.
This is also consistent with Campbell picture where the crossover from 
compact to ramified structure in the space of configurations is not sharp.

\section{Conclusions}
\label{conclusions}
In fully frustrated models we have numerically 
found an exponential relaxation above the
percolation temperature $T_p$, while for $T<T_p$ the long time tail 
of the relaxation functions can be fitted with a stretched exponential 
in both $2d$ \cite{ref_fierro} and $3d$ systems.

These results suggest that the percolation transition may play a role in the 
dynamical transition of frustrated systems without disorder. In particular, the 
role can be understood considering the physical meaning of the percolation 
transition in a generalization of the SG model (the $q$-state Potts SG 
\cite{FC_PRE}).
We suggest that the percolation transition marks the appearing 
of the ``large scale'' effects of frustration. Below the percolation
temperature, because of the presence of a spanning cluster, 
bond loops of any dimension may be closed, and therefore global effects of 
frustration are observed.

Note that in a previous paper \cite{ref_fierro} we also studied a model, the 
``locally frustrated bond percolation'', in which only frustrated loops 
whose length is equal to four were forbidden. 
The model exhibits the same critical properties as the random bond percolation,
showing that this kind of frustration is 
``too local'' to change the universality class of transition.
Similarly, the relaxation functions in the long time regime can always be 
fitted with an exponential function.

\section*{aknowledgement}
We would like to thank Chris Hanley for helpful comments.
This work was supported in part by the European TMR Network-Fractals
c.n.FMRXCT980183.
Simulation have been done on parallel Cray T3D at 
CINECA, Bologna.

\begin{table}
\begin{tabular}{|c|cc|}

$T$ & $\beta$ & $\tau^\beta$ \\
\hline
$5.0~$ & $0.99\pm 0.02$ & $0.16\pm 0.01$\\
$3.5~$ & $1.00\pm 0.02$ & $0.19\pm 0.01$\\
$2.5~$ & $0.98\pm 0.02$ & $0.24\pm 0.01$\\
$2.0~$ & $0.98\pm 0.02$ & $0.29\pm 0.01$\\
$1.7~$ & $0.94\pm 0.02$ & $0.44\pm 0.01$\\
$1.3~$ & $0.75\pm 0.02$ & $1.1 \pm 0.2$ \\
$0.7~$ & $0.51\pm 0.02$ & $2.8 \pm 0.2$ \\
\end{tabular}
\caption{Fit parameters for the autocorrelation functions $f(t)$ calculated 
with bond flip dynamics for the $q=2$-bond FF percolation model}
\label{tabella}
\end{table}

%
%
%
%%%%%%%%%%%%%%%%   REFERENCES  %%%%%%%%%%%%%%%%%%%%%%%%%%%%%%%%%%%%%%%%
%

%
%%%%%%%%%%%%%%%%%%%%%%%%%%%  FIGURES  %%%%%%%%%%%%%%%%%%%%%%%%%%%%%%
%
%%%%%%%%%%%%%%%%% fig. 1
%
%%%%%%%%%%%%%%% fig. 2a, 2b   
%
%
%%%%%%%%%%%%% fig. 3
%
%%%%%%%%%%%%%%%%% fig. 4
%%%%%%%%%%%%% fig. 5
%
%%%%%%%%%%%%%%%%% fig. 6
%
%%%%%%%%%%%%%%%%% fig. 7
%
%
%%%%%%%%%%%%%%%%% fig. 8
%
%
%%%%%%%%%%%%%%%%% fig. 9
%
%
%%%%%%%%%%%%%%%%% fig. 10

\end{document}